\documentclass[showpacs,manuscript,pre]{revtex4} 
\usepackage{amssymb}
\usepackage{amsmath}
\usepackage{amsthm}
\usepackage{graphicx}
\usepackage{dcolumn}
\usepackage{bm}

\newcommand{\ep}{\epsilon}

\newcommand{\obs}{{\cal O}}

\newtheorem{proposition}{Proposition}
\newtheorem{theorem}{Theorem} 
\theoremstyle{definition}

\theoremstyle{remark}

\begin{document}

\title{Stability criteria for $q$-expectation values}

\author{Rudolf Hanel$^{1}$ and  Stefan Thurner$^{1,2}$}

\affiliation{
$^1$ Complex Systems Research Group; HNO; Medical University of Vienna; 
W\"ahringer G\"urtel 18-20; A-1090; Austria \\
$^2$ Santa Fe Institute; 1399 Hyde Park Road; Santa Fe; NM 87501; USA\\
} 


\begin{abstract}
%
In statistical physics lately a specific kind of average, called the $q$-expectation value, 
has been extensively used in the context of $q$-generalized statistics dealing with distributions 
following power-laws. In this context $q$-expectation values appear naturally.
After it has been recently shown that this non-linear functional is instable, under a 
very strong notion of stability,
it is therefore of high interest to know sufficient conditions for when 
the results of $q$-expectations are robust under small variations of the underlying 
distribution function.  
We show that reasonable restrictions on the domain of 
admissible probability distributions restore uniform continuity for the 
$q$-expectation. Bounds on the size of admissible variations can be given.
The practical usefulness of the theorems for estimating the robustness of the $q$-expectation value
with respect to small variations is discussed.
\end{abstract}

\pacs{
05.20.-y, 
02.50.Cw, 
05.90.+m 
}

\maketitle
 
\section{Introduction}

In the context of generalizations of entropy-functionals generalized momenta occur naturally \cite{naudts_ent}, 
which, in the case of Tsallis $q$-statistics \cite{tsallis88} are commonly called escort distributions.
Aside from their necessity in several aspects of  $q$-statistics, expectation values under 
these escort distributions have been used to replace ordinary constraints in the maximum 
entropy principle \cite{tsallismendesplastino}.  Maximizing under these escort constraints  (also called $q$-constraints) 
via functional variations with respect to distributions $p$, the classical Tsallis entropy, 
$S_q= -\int p \ln_{2-q} p$, produces the famous $q$-exponential distributions, where
the the $q$-exponential function is defined as $\exp_q(x)=(1+(1-q)x)^{1/(1-q)}$. 
However, note that in general there is no need for $q$-constraints in the Tsallis formalism; 
the same $q$-exponential distributions can be derived under {\rm ordinary} constraints when Tsallis 
entropy is expressed in its dual form, $S_q= -\int p \ln_{q} p$, see \cite{ht1}.
The way generalized momenta still occur is when differential properties of ordinary expectation values
are considered \cite{naudts_ent}. For example, one may look at the $q$-exponential distribution 
$\exp_q(-\alpha-\beta \ep_i)$, where $\ep_i$ are discrete energy states, $\beta$ is the inverse temperature
and $\alpha$ is used for normalization, i.e. the normalization condition $1=\sum_i\exp_q(-\alpha-\beta \ep_i)$ holds.
The way $\alpha$ has to change with $\beta$, in this case, can be obtained by differentiating the normalization condition
with respect to $\beta$ and using $d\exp_q(x)/dx=\exp_q(x)^q$. Therefore,
\begin{equation}
\frac{d\alpha}{d\beta}=-\frac{\sum_i\exp_q(-\alpha-\beta \ep_i)^q\ep_i}{\sum_i\exp_q(-\alpha-\beta \ep_i)^q}\quad,  
\end{equation}
where the right side exactly corresponds to the $q$-expectation value
\begin{equation}
\langle\ep\rangle_q\equiv\frac{\sum_i p_i^q\ep_i}{\sum_ip_i^q}\quad,  
\end{equation}
when $p_i=\exp_q(-\alpha-\beta \ep_i)$ is the $q$-exponential distribution.
The distribution
\begin{equation}
P^{(q)}_i=\frac{p_i^q}{\sum_i p_i^q}\quad  
\end{equation}
usually is called the escort distribution of $p$.
One should note that, with respect to $p$, $q$-expectation values
\begin{equation}
\langle\obs\rangle_q\equiv=\sum_i P^{(q)}_i\obs_i\quad,  
\end{equation}
of some observables $\obs=\{\obs_i\}$ are non-linear functionals. 
In the entire paper we refer to the $q$-expectation value as a functional and will use the notation
$Q[p]\equiv\langle\obs\rangle_q$, to show its explicit dependency on $p$.
For all mathematical notions that will be used in this paper, like for instance {\emph equicontinuity},
{\emph uniform continuity} or {\emph Lebesgue decomposition}, we refer to standard
textbooks on functional analysis, like e.g. \cite{reedsimon1}.

It has to be noted that the question of continuity of functionals has been of some interest
lately, see e.g. \cite{lesche, fannes, abe2002, harremoes, naudts2004}.
Recently, it has been shown that, under small variations of the probability distribution, 
$q$-expectation values are instable in a certain sense \cite{abe}. 
It was concluded there that, due to this certain lack of stability, the usage of $q$-expectation values should be 
reconsidered in $q$-statistical physics. 
Therefore, it is important to ask whether this argument really disqualifies 
the usage of $q$-expectation values in general.  

The notion of stability used in \cite{abe} is closely related to stability in the sense of
Lesche \cite{lesche}. 
Let us write probabilities $p$ on $\mathbb{N}$ such that $\sum_i p_i =1$ and
the $||p||_1=\sum_i|p_i|$ is the $L_1(\mathbb{N})$-norm.
Probabilities on finite sets $i=1\dots W$ will simply be represented on $\mathbb{N}$ with
$p_i=0$ for all $i>W$, as in \cite{naudts2004}.
\\
\\
In \cite{abe} a 
functional $F[p]$ is called  
stable, if for all $\ep>0$ there exists a $\delta>0$, such that for all sequences of probabilities $p$
$\{p_W\}_{W=1}^{\infty}$ and $\{p'_W\}_{W=1}^{\infty}$, 
where $p_{W,i}=0$ and $p'_{W,i}=0$ for all $i>W$, it is true that
\begin{equation}
 \forall W[ \,\, ||p_W-p'_W||_1<\delta\,\,]   \,\,  \Rightarrow  \,\,  |F[p_W]-F[p'_W]|<\ep \quad.
\label{lesche} 
\end{equation}
Defining 
\begin{equation}
{\cal D}_0=\left\{ \{p_W\}_{W=1}^{\infty}\,\,|\,\, \forall W[\,\,||p_W||_1=1\,\,],\,\, \forall i>W[p_{W,i}=0] \right\}\quad,
\label{ddom0}
\end{equation}
the same definition, Eq. (\ref{lesche}) can be formulated shortly by calling a functional $F[p]$ stable if it
is uniformly equicontinuous on ${\cal D}_0$. 
To prove instability of a functional $F$ on ${\cal D}_0$ it is sufficient to find one example of a sequences $\{p_W\}_{W=1}^{\infty}\in{\cal D}_0$, such that uniformly equicontinuity of the functional $F$ is violated.
This is exactly what has been done in \cite{abe}. Two examples, one for $0<q<1$ and one for $1<q$, which originally
have been used by Lesche \cite{lesche} (for a detailed discussion see e.g. \cite{naudts2004}),
show that the $q$-expectation value $Q[p]$ is not uniformly equi-continuous on ${\cal D}_0$ and therefore
prove that $Q[p]$ is not stable. The recognition of such instabilities is important, since
they point at the fact that, under certain conditions or under certain circumstances,
it will be difficult to correctly estimate reliable values of $Q[p]$ (or any other functional, for instance entropies, that 
possesses an instability; see e.g. \cite{naudts2004}).

On the other hand, properties, like uniform continuity, are not simply properties of a functional
but are properties of a functional together with a domain of definition. Identification of the problematic regions,
in the domain of definition of the functional, therefore provides information on where on its domain the functional can 
be used without running into the instabilities the functional potentially possesses.
In the context of functions $\sqrt{x}$ may serve as an example. $\sqrt{x}$ is uniformly continuous on all 
intervals $[a,b]$, with $0<a<b<\infty$, but is not uniformly continuous on some interval $[0,b]$.
Uniform continuity fails when $0$ is an element of the considered domain of $\sqrt{x}$.
Similarly, we may ask whether reasonable domains ${\cal D}\subset{\cal D}_0$ can be found, such that
the functional $F$ is uniformly equi-continuous on ${\cal D}$, even though the functional is not uniformly equicontinuous
on ${\cal D}_0$. If such a ${\cal D}$ exists we can call $F$ stable on ${\cal D}$.
We will show in this paper that it is possible to find domains ${\cal D}$, such that the $q$-expectation value
$Q[p]$ is stable with respect to ${\cal D}$ as a functional. Moreover, the domains ${\cal D}$ are large enough to contain 
a large range of situations that usually are of physical interest. This will show that for this range of
practical situations the $q$-expectation value can safely be used and small variations of the distribution functions will not
lead to uncontrollable variations of the associated $q$-expectation values. 
The stability question in the case of $q$-expectation values 
is especially of interest as, for instance, it has been shown that a variety of correlated processes 
may lead to limit distributions that are extremely close to $q$-exponential functions
but are not $q$-exponential functions after all \cite{hilhorst07}. 
If in an effective theory experimental data should for practical means be misinterpreted in terms of 
$q$-exponential functions it therefore is crucial to know how reliable the predictions will be, given the experimental
uncertainty with respect to the underlying distribution. 

In order to understand the instability let us take a look at the two examples \cite{lesche,abe} 
violating uniform equi-continuity of the $q$-expectation value $Q[p]$, where
case (1) is associated with  $0<q<1$, and  case (2) with $1<q$. 
Specifically, in \cite{abe} the two cases are 
\\
case (1): $0<q<1$
\begin{equation}
p_i=\delta_{i\,1}\quad,\quad p_i'=\left(1-\frac{\delta}{2}\frac{W}{W-1}\right)p_i+\frac{\delta}{2}\frac{1}{W-1}
\label{case1}
\end{equation}
case (2): $1<q$
\begin{equation}
p_i=\frac{1}{W-1}\left(1-\delta_{i\,1}\right)\quad,\quad p_i'=\left(1-\frac{\delta}{2}\right)p_i+\frac{\delta}{2}\delta_{i\,1}\quad, 
\label{case2}
\end{equation}
where obviously $||p-p'||_1=\delta$, for any finite $W$. 
In the limit $W\to\infty$ both cases lead to 
$\lim_{W\to\infty}| Q[p]-Q[p']|=|\bar\obs-\obs_1|$, where $\bar\obs \equiv \lim_{W\to\infty}W^{-1}\sum_i \obs_i$, which
proves instability on ${\cal D}_0$ when $\obs$ and $K$ are chosen such that $|\bar \obs-\obs_1|>K>0$.
This is true, since in the limit there already exists a $W_0$ such that $| Q[p]-Q[p']|>K$ for all
$W>W_0$.

Though, this is not necessary for the validity of this proof one may note that 
the considered sequences of probabilities have a limit that is not a probability, i.e. 
the limit $W\to\infty$ and the $L_1(\mathbb N)$-norm do not commute. 
For instance, $\delta=\lim_{W\to\infty}\sum_{i=1}^W |p_i-p_i'|\neq\sum_{i=1}^\infty \lim_{W\to\infty} |p_i-p_i'|=\delta/2$, 
in both cases. This means that $p$ or $p'$ are in general not probabilities in the pointwise limit, 
e.g., in case $0<q<1$, one gets $\sum_i\lim_{W\to\infty}p'_i=(1-\delta/2)\neq 1$. 
The considered sequences of probabilities $\{p_W\}_{W=1}^{\infty}$
can easily be interpreted as a limit to distributions $\rho(x)$ on the continuous interval $x\in I\equiv[0,1]$
with $\int_I dx\,\rho=1$, where $dx$ is the usual Lebesgue measure on $[0,1]$.
We will analyze the stability problem within this continuous formulation. This is done
for two reasons. 
First, the continuity properties of $q$-expectation values
with respect to distribution functions $\rho(x)$, where $x\in[0,1]$, 
are of interest on their own, since power distributions are
not limited to discrete state spaces.
The second reason is that the discrete case is naturally embedded in the
continuous case, as demonstrated below. Propositions obtained in the continuous case can 
therefore be used to discuss continuity properties of the $q$-expectation value in the discrete case. 
In the continuous case we will denote the $q$-expectation with 
$\tilde Q[\rho]\equiv \int dx\, \rho^q(x)\obs(x)/\int dx\, \rho^q(x)$, where the observable
$\obs(x)$ now is suitable measurable function on $[0,1]$.

\subsection{The problem formulated for continuous distributions}\label{contlim}

The problem of the ill defined limit probabilities of the examples (1) and (2) is easily resolved by mapping
the discrete probabilities $\{p_{W,i}\}_{i=1}^W$ onto step functions $\rho_W(x)$, with $x\in[0,1]$ such that
$\rho_W(x)=W p_i$ for $x\in[(i-1)/W,i/W)$ (the last interval is chosen $[(W-1)/W,1]$). 
Therefore, for the usual Lebesgue measure $dx$ on $[0,1]$ it follows that $\int_0^1 dx \rho_W(x)=\sum_{i=1}^W\int_{(i-1)/W}^{i/W}\,\rho_W(x)=\sum_{i=1}^W p_{W,i}=1$, 
and the $L_1(\mathbb N)$-norm $||\rho-\rho'||_1=\int dx |\rho(x)-\rho'(x)|=\sum_i |p_i-p_i'|=||p-p'||_1 $.
Similarly, the discrete observable $\obs$ is mapped to a step function in an analogous way by identifying
$\obs_i=\obs(x)$ when $\in[(i-1)/W,i/W)$.
The discrete and the continuous $q$-expectation value therefore coincide since 
$\tilde Q[\rho]=\int dx\, \rho^q(x)\obs(x)/\int dx\, \rho^q(x)=\sum_i p_i^q\obs(x)/\sum_i p_i^q\equiv Q[p]$. 
In this way the limit $W\to\infty$ can be interpreted as the continuum limit of the step-functions $\rho$ and $\rho'$. 
These limits are well-defined probability distributions, and the $L1$-norm of the distributions and the $W\to\infty$  
limit commute.

In this continuum formulation the limit distributions of the families of distributions examples, case (1) and (2), that have violated uniform equi-continuity are given by
\\
case (1): $0<q<1$
\begin{equation}
\rho(x)=\delta(x)
\quad,\quad
\rho'(x)=(1-\frac{\delta}{2})\delta(x)+\frac{\delta}{2}\quad 
\label{cont1}
\end{equation}
case (2): $1<q$
\begin{equation}
\rho(x)=1
\quad,\quad 
\rho(x)'= 1-\frac{\delta}{2}+\frac{\delta}{2}\delta(x)\quad,
\label{cont2}
\end{equation}
where $\delta(x)$ is the usual delta function.
The result of \cite{abe}, that in the limit $W\to\infty$, $|Q[p]-Q[p']|=|\bar \obs -\obs_1|>0$ for $||p-p'||=\delta$, 
in the continuum  translates into 
that 
\begin{equation}
|\tilde Q[\rho]-\tilde Q[\rho']|=|\obs_1-\bar \obs |
\quad ,
\label{exactsol}
\end{equation} 
for $||\rho-\rho'||=\delta$, with $\obs_1=\obs(0)$ and $\bar \obs \equiv \int dx \obs(x)$.
Therefore, the first requirement we have to impose on ${\cal D}$ is, that the sequences
$\{p_W\}_{W=1}^{\infty}\in{\cal D}$ possess a continuum limit in $[0,1]$ with respect to the
$1$-norm on $L_1([0,1])$. Let us denote the set of all limit distributions produced by the sequences in  
${\cal D}$ with $\tilde{\cal D}$. If uniform equicontinuity of $Q[p]$ with respect to ${\cal D}$ has to hold
it is therefore necessary that $\tilde Q[\rho]$ is uniformly continuous on $\tilde{\cal D}$. 
This serves as starting point of the analysis.

The rest of the paper is organized as follows.
In section \ref{instcontlim} we present two theorems for the cases (1) $0<q<1$ and (2) $1<q$
that allow to analyze the continuity of $\tilde Q[\rho]$ around the distribution $\rho$.
The bounds given in the theorems are such that obvious definitions of the domain 
$\tilde{\cal D}$ of $\tilde Q[\rho]$ guarantees uniform continuity of the $q$-expectation 
value $\tilde Q[\rho]$ on these domains. An upper bound of admissible variations
on these domains is discussed which can be seen as a measure of overall robustness of
the $q$-expectation values on these domains, which may provide a practical mean to
check experimental situations for their robustness. 
In the discussion \ref{discussion} we will show how the theorems can be used in two examples. 
First, we will discuss there how the properties of $\tilde{\cal D}$ can be 
pulled back to a suitable ${\cal D}$ so that the $q$-expectation value $Q[p]$ becomes
uniform equicontinuous on ${\cal D}$.
Second, we will briefly discuss how the theorems can be used to analyze the continuity properties
of $\tilde Q[\rho]$ for distributions defined on the infinite interval $[0,\infty]$. 
This result allows to consider
a different subclass of ${\cal D}'\subset{\cal D}_0$ where
sequences of probabilities $p$ in ${\cal D}'\subset{\cal D}_0$,
possess a limit with respect to the $1$-norm on $L_1(\mathbb{N})$ and where the $1$-norm  on $L_1(\mathbb{N})$ and the limit $W\to\infty$ commute. 

\section{The instability in the general case}\label{instcontlim}
In the continuum the escort distribution reads  
$P^{(q)}(x)\equiv \frac{\rho(x)^q }{ \int dx' \rho(x')^q } $. 
The expectation value of a function $\obs(x)$ 
under this measure -- the $q$-expectation value -- is  then  
$\tilde Q[\rho] = \int dx  P^{(q)}(x)\obs(x)$. 
%
%
The total variation of $\tilde Q[\rho]$ therefore reads
\begin{equation}
\delta\tilde Q[\rho]=\tilde Q[\rho+\delta\rho]-\tilde Q[\rho]\quad.
\label{defunct}
\end{equation}
We can now analyze the two cases separately. 
The following proofs are carried out on the unit interval $I \in [0,1]$. This does not 
present a loss of generality, since the proofs can be extended to any bounded interval. 
For unbounded intervals,  especially relevant for $q>1$, the proofs get more involved and require 
to fix conditions that relate to specific boundedness conditions for the observable and decay properties 
of $\rho$, in order to keep $\tilde Q[\rho]$ a meaningful quantity as is briefly discussed in
section \ref{discussion}. 
%

\subsection{The case  $0<q<1$}

Looking at Equ. (\ref{cont1}) one can suppose that the uniform continuity property
of the $q$-expectation value $\tilde Q[\rho]$ is discontinuous for $\rho(x)=\delta(x)$
since is a pure point measure.
Due to Lebesgue decomposition for distributions each distribution $\rho$ can
be decomposed into a singular part $\rho_s$, that is defined on a set of Lebesgue measure zero,
and an absolute continuous part $\rho_c$.
We therefore assume that the distribution $\rho$ in the theorem is not purely singular, i.e.
it possesses an absolute continuous part $\rho_c$ with $\int_I dx \rho_c > 0$. 
Note that $||f||_p=(\int_I dx\,|f|^p)^{1/p}$ is the usual $p$-norm on $I=[0,1]$ and 
$||f||_\infty=\sup\{|f(x)|\,|\,x\in [0,1]\}$.

In order to prove Theorem (1), 
we have to establish  propositions \ref{prop1} - \ref{prop10},  see Appendix A.

\begin{theorem}
Let $0<q<1$. Let $\tilde Q[\rho]\equiv\langle\obs\rangle_q$ be the associated $q$-expectation 
value for the observable $\obs$. Let the distribution $0<\rho$ on $I=[0,1]$ have a non vanishing 
absolute continuous (non-singular) part. 
Let $G=\int_I dx\,\rho(x)^q$ and let $0<\tilde\delta^q=\mu G/4$ for $0<\mu<1$ and $\delta\rho$ be a 
variation of the distribution such that $\int_I dx |\delta\rho|=\delta\leq\tilde\delta$, and $0<\rho+\delta\rho$ 
is positive on $I$. 
Furthermore let $0<\obs$ be a strictly positive bounded observable on $I$, then there exists a constant 
$0<c<\infty$, such that
\begin{equation}
|\tilde Q[\rho]-\tilde Q[\rho+\delta\rho]|<c\delta^q\quad.
\end{equation}
Moreover $c\leq 4G^{-2}||\obs||_{\infty}(1+||\obs||_\infty||\obs^{-1}||_\infty)/(1-\mu)$.
\label{theorem1}
\end{theorem}
\begin{proof}
The requirement that $\rho$ is not purely singular is sufficient to 
guarantee that $0<\int_I dx \rho^q\obs$ is strictly positive.
Inversely, suppose $\rho(x)=\delta(x-x_0)$ is concentrated around one point 
$x_0$ and use the characteristic function $\Delta^{-1}\chi_{[-\frac{\Delta}{2}\,\frac{\Delta}{2}]}(x-x_0)$ as a $\delta$-sequence.
The characteristic function $\chi_{[a,b]}(x)=1$ for $x\in[a,b]$ and zero otherwise. 
It is straight forward to see 
that $\int_{D_+} dx\, \rho(x)^q=\int_0^\Delta dx \Delta^{-q}=\Delta^{1-q}\to 0$, for $\Delta\to 0$. 
This can not happen if $\rho$ has a non vanishing  absolutely continuous part.  

Note that
\begin{equation}
|\tilde Q[\rho]-\tilde Q[\rho+\delta\rho]|=|\tilde Q[\rho]||1-\tilde Q[\rho+\delta\rho]\tilde Q[\rho]^{-1}|\quad.
\end{equation}
Using the H\"older-inequality one finds $\int_I dx\, \rho(x)^q\obs(x)\leq ||\obs||_{1/(1-q)}\leq||\obs||_\infty$.
Consequently $|\tilde Q[\rho]|<||\obs||_\infty/G$. 
Furthermore, note that $\int_I dx\, \rho(x)^q\obs(x)\geq G/||\obs^{-1}||_\infty$.
Propositions (\ref{prop1}-\ref{prop10}) imply that
\begin{equation}
\left(1-\frac{\tilde C_2\delta^q}{G-\tilde C_3 \tilde\delta^q}\right)\left(1-\frac{C_3\delta^q}{\int_I dx\,\rho^q\obs}\right)
\leq
\frac{\tilde Q[\rho+\delta\rho]}{\tilde Q[\rho]}
\leq
\left(1+\frac{C_2\delta^q}{\int_I dx\,\rho^q\obs}\right)\left(1+\frac{\tilde C_3 \delta^q}{G-\tilde C_3 \tilde\delta^q}\right) 
\end{equation}
Setting the constants to their upper bounds, i.e. $C_2\to 4||\obs||_\infty$, $C_3\to 4||\obs||_\infty$, 
$\tilde C_2\to 4$, $\tilde C_3\to 4$, and evaluating the terms of the left and the right side gives 
$(1-a_1\delta^q+a_2 \delta^{2q})
\leq
\tilde Q[\rho+\delta\rho]/\tilde Q[\rho]
\leq (1+b_1\delta^q+b_2 \delta^{2q})
$ and the resulting constants $a_1$, $a_2$, $b_1$, and $b_2$ are all positive.
On the left side we note that $1-a_1\delta^q\leq1<1-a_1\delta^q+a_2 \delta^{2q}$ and on the right side
$1+b_1\delta^q+b_2 \delta^{2q}<1+b_1\delta^q+b_2 \tilde\delta^q\delta^q$. Furthermore,
$a_1<b_1+b_2\tilde\delta^q$. This allows to give an upper bound for $c$ given by 
$c=||\obs||_\infty(b_1+b_2\tilde\delta^q)/G$.
Moreover, $b_1=(4/(1-\mu)+4||\obs||_\infty||\obs^{-1}||_\infty)/G$ and 
$b_2\tilde\delta^q=4G^{-1}||\obs||_\infty||\obs^{-1}||_\infty\mu/(1-\mu)$
which completes the proof.
\end{proof}

The theorem (together with its associated propositions) states that for strictly positive bounded observables $q$-expectation 
values are continuous for non purely singular $\rho$, i.e. the absolute continuous part of $\rho$ is non vanishing. 
Clearly uniform continuity of the $q$-expectation value can not be established on all of $L_1([0,1])$.
However, it follows from Theorem (1) that on any domain 
\begin{equation}
\tilde{\cal D}^{(1)}_{B,r}=\{\rho| 0<\rho\in L_1([0,1]), ||\rho||_1=1, 0\leq r\leq\int_I dx\,\rho(x)^q\leq B\}
\label{dom1}
\end{equation}
the $q$-expectation value $\tilde Q[\rho]$ is uniformly continuous.
The lower bound $r$ on $\int_I dx\,\rho(x)^q$ is required in order to exclude distributions
with purely singular measure\footnote{Since $0<q<1$ a more restrictive way to exclude purely singular distributions is to
require boundedness of the distributions $\rho$ from above.}. 
The constant $c$ in general is depending on $\rho$ since 
$G=\int_I dx\,\rho(x)^q$. However due to the common lower bound $r$ it follows that $G\geq r$ on all $\rho\in\tilde{\cal D}_{B,r}$.
Therefore, choosing $c = 4r^{-2}||\obs||_{\infty}(1+||\obs||_\infty||\obs^{-1}||_\infty)/(1-\mu)$ is a sufficiently large
on all of $\rho\in\tilde{\cal D}_{B,r}$ and
$c$ does not depend on the particular choice of $\rho\in\tilde{\cal D}_{B,r}$ any more.
Consequently, 
uniform continuity of the $q$-expectation value $\tilde Q[\rho]$ 
is established on any domain $\rho\in\tilde{\cal D}_{B,r}$ .
Further, since $\tilde\delta=(\mu G/4)^{1/q}$ is an upper bound on the $L_1$-norm $\delta=||\delta \rho||_1$ of variations 
$\delta \rho=\rho-\rho'$, guaranteeing the validity of
$|\tilde Q[\rho]-\tilde Q[\rho']|\leq c \delta$.
Most likely these bound can be improved. Yet, $\tilde\delta$ can be seen as a measure of robustness
of the $q$-expectation value $\tilde Q[\rho]$ on $\tilde{\cal D}_{B,r}$.
To make $\tilde\delta$ independent of the choice of $\rho$ one has to set
$\tilde\delta=(\mu B/4)^{1/q}$. 
It has to be noted that the upper bound $\tilde\delta$ 
decreases with increasing $B$ like $0<\tilde\delta^q = \mu B^{q-1}/4$ and therefore
robustness under variations will in general decrease with increasing $B$.

We want to remark that the condition of strict positivity
of the observable, we have required as a condition in the theorem, can be relaxed to observables that are bounded from below
by some constant $L$, i.e. $\obs\geq L>-\infty$. If this is the case, one can look at the observable 
$\obs_L=\obs-L+1$, which is strictly positive and $||\obs_L^{-1}||_\infty=1$. 
Since for the $q$-expectation value it is true that $\rangle 1 \langle_q=1$ for any admissible distribution
it is also true that $\rangle \obs_L \langle_q=\rangle \obs \langle_q-L+1$.
The results therefore relax to bounded observables, i.e $||\obs||_\infty<\infty$.
By shifting $\obs$ to $\obs_L$ we can make the substitutions in the bounds 
$||\obs_L||_\infty\to 2||\obs||_\infty+1$ and $||\obs_L^{-1}||_\infty\to 1$

\subsection{The case $1<q$}

In contrast to the $0<q<1$ case,  the instability in the $1<q$ case is not 
caused by purely singular distributions $\rho$, but due to the variation 
$\delta\rho$ having a non vanishing singular part.
In order to prove Theorem (2), 
we have to establish  propositions \ref{prop12} - \ref{prop17},  see Appendix B.

\begin{theorem}
Let $q>1$ and let $m>0$ be an arbitrary but fixed constant.
Let $0<\rho$ be a probability distribution on $I=[0\,1]$.
Let $\delta \rho$ be variations of $\rho$, i.e. $\rho+\delta\rho>0$. 
Let $\tilde Q[\rho]=\langle\obs\rangle_q$ be the $q$-expectation value
and let $0<\obs$ be a strictly positive bounded observable on $I$.
Let $B>0$ be an arbitrary but fixed constant.
Let the variations $\delta\rho$ be uniformly bounded in the $m$-norm, such that $||\delta\rho||_m<B$.
Further let $||\delta\rho||_1=\delta$.
Let $\tilde\delta$ be an upper bound for the size of the variations $\delta$ such that
$(2^{1/q}-1)^{q/\gamma}B^{(\gamma-q)/\gamma}\left(\min\left(1,||\obs||_\infty||\obs^{-1}||_\infty\right)\right)^{-1/\gamma} \geq \tilde\delta > 0$, 
where $\gamma=(m-q)/(m-1)$.
Then, there exists a constant $0<R<\infty$, such that
\begin{equation}
|\tilde Q[\rho]-\tilde Q[\rho+\delta\rho]|<R\delta^{\gamma/q}\quad,
\end{equation}
and $R$ does not depend on the choice of $\rho$.
\label{theorem2}
\end{theorem}

\begin{proof}
This result follows directly from propositions (\ref{prop12}-\ref{prop17}) from Appendix B, and by noting that
\begin{equation}
\frac{1-R_2 \delta^{\gamma/q}}{1+\tilde R_2 \delta^{\gamma/q}}
\leq
\frac{\tilde Q[\rho+\delta\rho]}{\tilde Q[\rho]}
\leq
\frac{1+R_2 \delta^{\gamma/q}}{1-\tilde R_2 \delta^{\gamma/q}}
\quad.
\end{equation}
Proposition \ref{prop17} tells us that $1/(1-\tilde R_2 \delta^{\gamma/q})\leq 
1+R_3\delta^{\gamma/q}$. Moreover $1/(1+\tilde R_2 \delta^{\gamma/q})\geq 1-\tilde R_2 \delta^{\gamma/q}$.
Note that $R_3>\tilde R_2$. Since $||\obs^{-1}||_\infty^{-1}\leq \tilde Q[\rho]\leq||\obs||_\infty$
choosing $R=R_2R_3\max\{||\obs^{-1}||_\infty^{-1},||\obs||_\infty\}$ is sufficient.
Noting that both $R_2$ and $R_3$ are not depending on the particular choice of
$\rho$ completes the proof.
\end{proof}

The theorem (together with its associated propositions) states that, for strictly positive bounded observables, $q$-expectation 
values are continuous for any $\rho$, as long as the variation $\delta \rho=\rho'-\rho$ is bounded 
in some $m$-norm with $m>q$.
By considering domains 
\begin{equation}
\tilde{\cal D}^{(2)}_{B,m}=\{\rho| \rho\in L_1([0,1])\bigcap L_m([0,1]), ||\rho||_m\leq B\}
\label{dom2}
\end{equation}
for case (2), i.e. $1<q$, automatically any admissible variation $||\delta\rho||_m<B$ and 
the constant $R$ is not depending on the particular choice of admissible variation with respect to the 
domain $\tilde{\cal D}_{B,m}$ any more.
This proves that the $q$-expectation value $\tilde Q[\rho]$ is uniformly continuous on any $\tilde{\cal D}_{B,m}$
with $m>q$.
Again, it has to be noted that $\tilde \delta\propto B^{(\gamma-q)/\gamma}$.
Since $0<\gamma<1$ and $q>1$ it follows that $(\gamma-q)/\gamma<0$ and
$\tilde \delta$ decreases as $B$ increases. 
Measuring robustness in $\tilde\delta$ again shows that robustness of the $q$-expectation value 
with respect to small variations decreases with enlarging the domain of definition as
expected.

\section{Discussion}\label{discussion}

We will now demonstrate the practicability of the two theorems
by discussing two applications of the theorems.
The first application is to understand when uniform equicontinuity
of families of sequences of probabilities can be expected.
The second application is to extend the conditions for uniform continuity
of the $q$-expectation value from the case where the distributions have
compact support, i.e. $[0,1]$, to the case where distributions have an unbounded support
$[0,\infty]$.  

First, we turn to the question of uniform equicontinuity the $q$-expectation value.
We have shown in section \ref{instcontlim} that
$q$-expectation values are uniformly continuous for domains that in case (1) $0<q<1$ have been
specified in Eq. (\ref{dom1}) and in case (2) $1<q$ in Eq. (\ref{dom2}).
These results allow to establish equicontinity properties of $q$-expectation values
for sequences of probabilities $\{p_W\}_{W=1}^{\infty}\in{\cal D}\subset{\cal D}_0$, specified in Eq. (\ref{ddom0}).

To make the contact with the continuum results, we have to impose that the limit of the sequences 
of probabilities in ${\cal D}$ exists as continuum limits in the $L_1([0,1])$-norm,
i.e. in terms of step functions $\rho_W$ representing $p_W$, as described in section \ref{contlim}. 
The span of these limits has to coincide with the domain $\tilde{{\cal D}}$. 
This can be achieved when all the distributions $\rho_W\in\tilde{{\cal D}}$. 
In case (1) the conditions defining $\tilde{\cal D}_{B,r}$, for some $0<r<B$, translate into the requirement that
\begin{equation}
r\leq  W^{q-1} \sum_i p_{W,i}^q \leq B\quad.
\end{equation}
In case (2) the conditions defining $\tilde{\cal D}_{B,m}$, for some $m>q>1$ and some $B>0$, translate into the requirement that
\begin{equation}
||p_W||_m \leq B W^{\frac{1-m}{m}}\quad.
\end{equation}
By imposing these conditions on the domain of sequences ${\cal D}$, uniformly equicontinuity
of the $q$-expectation value, with respect to ${\cal D}$, can be established for both cases (1) and (2).
Consequently, $q$-expectation values can be called \emph{robust} or stable with respect to the 
specified domains ${\cal D}$.

We discuss a second application of the theorems, to establish
criteria for specifying subsets of probability distributions $\rho\in L_1([0,\infty])$
such that the $q$-expectation value again is uniformly continuous on this domain.
Again, one can use the results of section \ref{instcontlim} as a starting point of the discussion
and proceed as follows. 
 
Choose a suitable differentiable monotonous functions,
$g: [0,\infty]\mapsto [0,1]$.  Let $g'$ denote the derivative of $g$ and
$g^{-1}$ the inverse function of $g$.
Therefore, $g$ maps the distribution function $\rho$, defined on $[0,\infty]$, to a
distribution function
$\tilde \rho(y)=\rho(g^{-1}(y))g'(g^{-1}(y))^{-1}$ on $[0,1]$.
Similarly, the observable function $\obs$ on $[0,\infty]$ gets mapped
to $\tilde\obs(y)=\obs(g^{-1}(y))g'(g^{-1}(y))^{q-1}$. 
Applying the conditions used for the theorems 1 and 2 and 
characterizing domains where the $q$-expectation value on $[0,1]$ is
uniformly continuous poses restrictions on 
the transformed distributions $\tilde\rho$ and 
the transformed observable $\tilde\obs$.
These restrictions can now be pulled back to the distribution
$\rho$ and the observable $\obs$ on $[0,\infty]$.
For specific problems different choices of $g$ may be considered.
It is instructive to look at an explicit example.
Consider $\bar q$-exponential distributions  
$\rho(x) \propto e_{\bar q }(-\beta x)\equiv [1-(1-\bar q)\beta x]^{\frac{1}{1-\bar q } }$
for $\bar q \ge 1$ and some inverse temperature $\beta$.
Assume that we wish to measure the first $N$ moments 
under the $q$-expectation,
\begin{equation}
\langle x^n \rangle_q\equiv \frac{\int dx [\rho (x)]^q x^n}{\int dx [\rho (x)]^q }\quad,
\label{ho}
\end{equation}
in a reliable way (i.e. $n\leq N$).
Assume $q>1$ and consider $\tilde{\cal D}_{B,\infty}$ as the admissible domain of distributions on $[0,1]$ (i.e $m=\infty$). 
It follows that $B>||g'(x)\rho(x)||_\infty$.
Choose $g(x)=1-1/(1+x)^\phi$ for some $\phi>0$. Consequently, $g'(x)=\phi (1+x)^{-\phi-1}$. 
The boundedness condition for the observables immediately  requires 
$\phi>N/(q-1)-1$ and the decay property for the distributions implies 
$\bar q<1+1/(\phi+1)$.
Inversely, this means that for specific distributions $\rho$ on $[0,\infty]$ 
it is possible to design domains around these specific $\rho$ where the $q$-expectation value is
uniformly continuous.
Again, the discrete case of probabilities $p$ on $\mathbb{N}_+$ into $[0,\infty]$
can be embedded in the continuous case $[0,\infty]$ using step-functions $\rho_p(x)=p_i$ for $x\in[i-1, i)$.
Domains $\tilde{\cal D}'$ of uniform continuity of the $q$-expectation value of distributions on 
$[0,\infty]$ can be pulled back to domains of ${\cal D}'$ such that the $q$-expectation value is uniformly equi-continuous 
on ${\cal D}'$.

\section{Conclusion}\label{conclusion}
To summarize, we have shown that reasonable restrictions on the domain of 
admissible probability distributions restore uniform continuity for the 
$q$-expectation on this domain.
Bounds on the size of admissible variations have been given that allow
to estimate the overall robustness of the $q$-expectation under small
variations.
The practical usefulness of the theorems for estimating the robustness of the $q$-expectation value
with respect to small variations has been discussed.
%
\\
\\ 
This work was  supported by Austrian Science Fund FWF Projects P17621 and P19132.

\section*{Apendix A}

This appendix contains the propositions for the proof of theorem (1), the case $0<q<1$.

\begin{proposition}
Let $D\subset I\equiv[0,1]$ and $0<\obs$ be a bounded positive function on $I$. Further,
let $\delta\rho$ be a function on $I$ such that $\int_I dx\,|\delta\rho|=\delta$, then there exists a constant $0<C_1<\infty$, such that
$\int_D dx\, |\delta\rho(x)|^q \obs(x)\leq C_1\delta^q$. Furthermore, $C_1<||\obs||_\infty|D|^{1-q}$, where
$|D|=\int_D dx$.
\label{prop1}
\end{proposition}

\begin{proof}
Using the H\"older-inequality find, 
$\int_D dx |\delta\rho|^q\obs\leq\left(\int_D dx\,\delta\rho\right)^q\left(\int_D dx\,|\obs|^{\frac{1}{1-q}}\right)^{1-q}$.
Setting $C_1=\left(\int_D dx\,|\obs|^{\frac{1}{1-q}}\right)^{1-q}<(||\obs^{\frac{1}{1-q}}||_\infty|D|)^{1-q}=||\obs||_\infty|D|^{1-q}$, and noting that
$\left(\int_D dx\,|\delta\rho|\right)^q\leq\left(\int_I dx\,|\delta\rho|\right)^q=\delta^q$, completes the proof.
\end{proof}

\begin{proposition}
Let $D_0=\{x|\rho(x)=0\}$ and $D_+=\{x|\rho(x)>0\}$ and $r=\int_{D_+} dx \equiv |D_+|=1-|D_0|$, then
$\int_{D_0} dx\, |\delta\rho|^q\leq \delta^q(1-r)^{1-q}$, and
$\int_{D_+} dx\, |\delta\rho|^q\leq \delta^q r^{1-q}$.
\label{prop2}
\end{proposition}

\begin{proof}
Set $\obs=1$ in proposition (\ref{prop1}) and use $r=|D_+|$.
\end{proof}

\begin{proposition}
Let $0<\rho$ be a non-singular probability distribution on $I=[0,1]$ and $\delta\rho$ be a variation of the distribution
such that $\int_I dx |\delta\rho|=\delta$, and $0<\rho+\delta\rho$ is positive on $I$. Further, let $0<\obs$ be a positive bounded
observable on $I$, then there exists a constant $0<C_2<\infty$ 
such that
\begin{equation}
\int_I dx\,(\rho+\delta\rho)^q\obs\leq \left(\int_I dx\,\rho^q\obs\right) + C_2\delta^q \quad ,
\end{equation}
and  $C_2=||\obs||_1+2q||\obs||_\infty+C_1 \leq 4||\obs||_\infty$.
\label{prop3}
\end{proposition}

\begin{proof}
Let $D_+^-=\{x|0<\rho(x)\leq\delta\}$ and $D_+^+=\{x|\delta<\rho(x)\}$, then
$\int_I dx (\rho+\delta\rho)^q\obs=\int_{D_+^-} dx (\rho+\delta\rho)^q\obs + \int_{D_+^+} dx (\rho+\delta\rho)^q\obs
+\int_{D_0} dx \delta\rho^q\obs$.
Since a power of $q<1$ is concave the first term leads to 
$\int_{D_+^-} dx\, (\rho+\delta\rho)^q\obs 
\leq 
\int_{D_+^-} dx\,(\delta^q+q\delta^{q-1}(\delta\rho+\rho-\delta))\obs
\leq
\delta^q \left(\int_{D_+^-} dx\obs\right)+q\delta^{q-1}\left(\int_{D_+^-} dx\,|\delta\rho|\obs\right)
\leq
\delta^q(||\obs||_1+q||\obs||_\infty)$.
Similarly, the second term leads to
$\int_{D_+^+} dx (\rho+\delta\rho)^q\obs 
\leq 
\int_{D_+^+} dx (\rho^q +q\delta^{q-1}|\delta\rho|)\obs
\leq
\int_I dx \rho^q\obs + q\delta^q||\obs||_\infty$. The third term, that corresponds to the part of the domain where
$\rho(x)=0$, is estimated by proposition (\ref{prop1}).
Adding all three contributions together leads to the result. 
\end{proof}

\begin{proposition}
Let $0<\rho$ be a non-singular probability distribution on $I=[0,1]$ and $\delta\rho$ a variation of the distribution
such that, $\int_I dx |\delta\rho|=\delta$, and $0<\rho+\delta\rho$ is positive on $I$, then there exists a constant 
$0<\tilde C_2<\infty$ 
such that
\begin{equation}
\int_I dx\,(\rho+\delta\rho)^q\leq \left(\int_I dx\,\rho^q\right) + \tilde C_2\delta^q \quad ,
\end{equation}
with $\tilde C_2=1+2q+(1-r)^{1-q} \leq 4$.
\label{prop4}
\end{proposition}

\begin{proof}
Use proposition (\ref{prop3}) and set $\obs=1$ to find $||\obs||_1=1$ and $||\obs||_\infty=1$.
\end{proof}

\begin{proposition}
Under the same conditions as in proposition (\ref{prop3}) find that
\begin{equation}
\left(\int_I dx\,\rho^q\obs\right) - C_3\delta^q 
\leq
\int_I dx\,(\rho+\delta\rho)^q\obs \quad , 
\end{equation}
with $C_3=||\obs||_1 +(2q+1)||\obs||_\infty<4||\obs||_\infty$.
\label{prop7}
\end{proposition}

\begin{proof}
Use proposition (\ref{prop3}) with $\rho=(\rho+\delta\rho)-\delta\rho$, 
i.e. substitute $\rho'=\rho+\delta\rho$ and $\delta\rho'=-\delta\rho$ and adapt $D_0$ and $D^\pm_+$ to $\rho'$ accordingly.
Due to the substitution $\rho\to\rho'$ the value of $r=|D_+|$ can not be assumed to remain invariant. 
Choosing the worst possible case, $r=0$, leads to the result.
\end{proof}

\begin{proposition}
Under the same conditions as in  proposition (\ref{prop3}) find that
\begin{equation}
\left(\int_I dx\,\rho^q\right) - \tilde C_3\delta^q 
\leq
\int_I dx\,(\rho+\delta\rho)^q \quad , 
\end{equation}
with $\tilde C_3=2(1+q)<4$.
\label{prop8}
\end{proposition}

\begin{proof}
Use proposition (\ref{prop7}) and set $\obs=1$.
\end{proof}

\begin{proposition}
Let $G=\int_I dx\,\rho(x)^q$ and let $0<\tilde\delta^q=\mu G/4$ for $0<\mu<1$.
Under the same conditions as in proposition (\ref{prop3}), it follows that for all 
$0<\delta<\tilde\delta$
\begin{equation}
\frac{\int_I dx \rho^q\obs}{\int_I dx (\rho+\delta\rho)^q\obs}-1\leq \frac{C_3\delta^q}{\int_I dx \rho^q\obs-C_3\tilde\delta^q}\quad . 
\end{equation}
\label{prop9}
\end{proposition}

\begin{proof}
Use proposition (\ref{prop7}) to get
$\int_I dx \rho^q\obs/\int_I dx (\rho+\delta\rho)^q\obs\leq 1+C_3\delta^q/\int_I dx (\rho+\delta\rho)^q\obs$.
Use proposition (\ref{prop7}) again on the right hand side to estimate $\int_I dx (\rho+\delta\rho)^q\obs$
from below and take the minimal admissible value of this estimate by setting $\delta^q$ to $\tilde\delta^q$.
\end{proof}

\begin{proposition}
Let $G=\int_I dx\,\rho(x)^q$ and let $0<\tilde\delta^q=\mu G/4$ for $0<\mu<1$.
Under the same conditions as in proposition (\ref{prop3}) it follows that
for all $0<\delta<\tilde \delta$,
\begin{equation}
\frac{\int_I dx \rho^q}{\int_I dx (\rho+\delta\rho)^q}-1\leq \frac{\tilde C_3\delta^q}{G-\tilde C_3\tilde\delta^q}\quad . 
\end{equation}
\label{prop10}
\end{proposition}

\begin{proof}
Repeat the proof of proposition (\ref{prop9}) for $\obs=1$, i.e. by using 
proposition (\ref{prop8}) instead of proposition (\ref{prop7}).
\end{proof}


\section*{Apendix B}

This appendix contains the propositions for the proof of theorem (2), the case $1<q$.  
Since $q>0$, the $q$-norm $||f||_q=(\int_I dx\,|f(x)|^q)^{1/q}$ is the usual $L_q$ norm. 

\begin{proposition}
Let $B>0$ be a arbitrary positive constant. Let $\delta\rho$ be functions on $I=[0, 1]$ that are uniformly bounded for 
some $m$-norm, i.e. $||\delta\rho||_m<B$, where $m>q$. Further let $||\delta\rho||_1=\delta$. 
Let $0<\obs$ be a positive bounded observable on $I$, 
then there exists a constant $0<R_1<\infty$, such that
\begin{equation}
\int_I dx\,|\delta\rho|^q\obs \leq R_1\delta^\gamma \quad ,
\end{equation}
where $\gamma=(m-q)/(m-1)\leq 1$ and 
and $R_1=B^{q-\gamma}||\obs||_\infty$.
\label{prop12}
\end{proposition}

\begin{proof}
Let $\gamma$ be a constant $0<\gamma\leq1$.
$\int_I dx\,|\delta\rho|^q\obs \leq ||\obs||_\infty||\,|\delta\rho|^\gamma|\delta\rho|^{q-\gamma}||_1
\leq||\obs||_\infty||\,|\delta\rho|^\gamma||_{1/\gamma}||\,|\delta\rho|^{q-\gamma}||_{1/(1-\gamma)}$ 
using H\"older's inequality.
Now choosing $a$ such that $m=(q-\gamma)/(1-\gamma)$ and noting that this means 
$\gamma=(m-q)/(m-1)$, i.e. 
$q-\gamma=(q-1)m/(m-1)$, we get
$||\,|\delta\rho|^\gamma||_{1/\gamma}||\,|\delta\rho|^{q-\gamma}||_{1/(1-\gamma)}=(||\delta\rho||_1)^\gamma
(||\delta\rho||_m)^{q-\gamma}\leq \delta^\gamma B^{q-\gamma}$.
\end{proof}

\begin{proposition}
Let $0<\rho$ be a probability distribution on $I=[0,1]$, i.e. $||\rho||_1=1$, with finite 
$q$-norm, i.e. $||\rho||_q<\infty$.
Let $\delta\rho$ be a variation of $\rho$, i.e. $0<\rho+\delta\rho$ is positive on $I$,
that has the properties specified in proposition (\ref{prop12}).
Further let $0<\tilde\delta$ be some positive constant and $||\delta\rho||_1=\delta\leq\tilde\delta$
and let $\obs>0$ be a strictly positive bounded observable then   
there exists a constant $0<R_2<\infty$, 
such that
\begin{equation}
\int_I dx\,(\rho+\delta\rho)^q\obs \leq  \left(1 +R_2\delta^{\gamma/q}\right)\int_I dx\,\rho^q\obs\quad.
\end{equation}
\label{prop13}
\end{proposition}

\begin{proof}
Since $1<q$ we first use the Minkowsky inequality and then proposition (\ref{prop12}) to get
$\int_I dx\,(\rho+\delta\rho)^q\obs 
\leq 
((\int_I dx\,\rho^q\obs)^{1/q}+(\int_I dx\,|\delta\rho|^q\obs)^{1/q})^q
\leq
((\int_I dx\,\rho^q\obs)^{1/q}+(R_1\delta^\gamma)^{1/q})^q
\leq
\int_I dx\,\rho^q\obs(1+(R_1\delta^\gamma/\int_I dx\,\rho^q\obs)^{1/q})^q
$
Now we note that the minimum for $\int_I dx\,\rho^q\obs$ can be obtaind for 
$\rho^{q-1}\propto\obs^{-1}$ and it follows that 
$\int_I dx\,\rho^q\obs\geq(||\obs^{-1}||_\infty)^{-1}$. 
Therefore 
$\int_I dx\,(\rho+\delta\rho)^q\obs 
\leq 
\int_I dx\,\rho^q\obs(1+z\delta^{\gamma/q})^q
$
where $z=(R_1||\obs^{-1}||_\infty)^{1/q}$.
Now we note that since $q>1$ for all $\delta<\tilde\delta$
it holds that 
$(1+z\delta^{\gamma/q})^q\leq 1+R_2\delta^{\gamma/q}$
with
$R_2=((1+z\tilde\delta^{\gamma/q})^q-1)/\tilde\delta^{\gamma/q}$.
\end{proof}

\begin{proposition}
Under the same conditions as in proposition (\ref{prop13}), there exists a constant $0<\tilde R_2<\infty$, such that
\begin{equation}
\int_I dx\,(\rho+\delta\rho)^q \leq  \left(1 +\tilde R_2\delta^{\gamma/q}\right)\int_I dx\,\rho^q\quad.
\end{equation}
\label{prop14}
\end{proposition}

\begin{proof}
Use proposition (\ref{prop13}) and set $\obs=1$.
\end{proof}

\begin{proposition}
Under the same conditions as in proposition (\ref{prop13}), and
for $0<\tilde\delta$ chosen small enough it holds that $0<\left(1-R_2\tilde\delta^{\gamma/q}\right)$
and
\begin{equation}
 \left(1-R_2\delta^{\gamma/q}\right)\int_I dx\,\rho^q\obs \leq \int_I dx\,(\rho+\delta\rho)^q\obs \quad.
\end{equation}

\label{prop15}
\end{proposition}

\begin{proof}
Use proposition (\ref{prop13}) with $\rho'=\rho+\delta\rho$ and $\delta\rho'=-\delta\rho$ to get
$\int_I dx\,\rho^q\obs\quad \leq  \left(1 +R_2\delta^{\gamma/q}\right)\int_I dx\,(\rho+\delta\rho)^q\obs$.
Then, divide this result by $\left(1 +R_2\delta^{\gamma/q}\right)$ and note that 
$1/(1+x)>1-x$ to get the result. In order for 
$0\leq\left(1-R_2\tilde\delta^{\gamma/q}\right)$ to hold simple calculation show that
this can be guaranteed by choosing $\tilde\delta$ small enough, i.e. 
$(2^{1/q}-1)^{q/\gamma}(B^{q-\gamma}||\obs||_\infty||\obs^{-1}||_\infty)^{-1/\gamma} > \tilde\delta$.
\end{proof}

\begin{proposition}
Under the same conditions as in proposition (\ref{prop13}), there exists a constant $0<\tilde R_3<\infty$, such that
\begin{equation}
 \left(1-\tilde R_2\delta^{\gamma/q}\right)\int_I dx\,\rho^q \leq \int_I dx\,(\rho+\delta\rho)^q \quad.
\end{equation}
\label{prop16}
\end{proposition}

\begin{proof}
Use proposition (\ref{prop15}) and set $\obs=1$.
\end{proof}

\begin{proposition}
Under the same conditions as in proposition (\ref{prop13}) and $0<\delta\leq\tilde\delta$, 
there exists a constant $0<R_3<\infty$, such that
\begin{equation}
\frac{\int_I dx\,\rho^q}{\int_I dx\,(\rho+\delta\rho)^q}-1 
\leq 
1+R_3\delta^{\gamma/q}\quad.
\end{equation}
\label{prop17}
\end{proposition}

\begin{proof}
Use proposition (\ref{prop16}) to get 
$\int_I dx\,\rho^q/\int_I dx\,(\rho+\delta\rho)^q\leq 
1/(1-\tilde R_2\delta^{\gamma/q})\leq
1+R_3\delta^{\gamma/q}$
with
$R_3=\left(1/(1-\tilde R_2\tilde\delta^{\gamma/q})-1\right)\tilde\delta^{-\gamma/q}
=\tilde R_2/(1-\tilde R_2\tilde\delta^{\gamma/q})$.
This completes the proof.
\end{proof}

\end{document}